\documentstyle[12pt,aaspp4,flushrt]{article}
\def\etal{{et al.}}
\def\name{SDSS 1624+00}
\begin{document}

\title{The Discovery of a Field Methane Dwarf from Sloan
Digital Sky Survey Commissioning Data\footnote{Based on observations obtained with the
Sloan Digital Sky Survey and the Apache Point Observatory
3.5-meter telescope, which are owned and operated by the Astrophysical
Research Consortium, and with the United Kingdom Infrared Telescope.}}
\author{
Michael A. Strauss\altaffilmark{\ref{Princeton}},
Xiaohui Fan\altaffilmark{\ref{Princeton}},
James E. Gunn\altaffilmark{\ref{Princeton}}, 
S. K. Leggett\altaffilmark{\ref{UKIRT}},
T.R. Geballe\altaffilmark{\ref{Gemini}},
Jeffrey R. Pier\altaffilmark{\ref{Flagstaff}},
Robert H. Lupton\altaffilmark{\ref{Princeton}}, 
G. R. Knapp\altaffilmark{\ref{Princeton}},
James Annis\altaffilmark{\ref{Fermilab}},
J. Brinkmann\altaffilmark{\ref{APO}}, 
J. H. Crocker\altaffilmark{\ref{JHU}},
Istv\'an Csabai\altaffilmark{\ref{JHU},\ref{Eotvos}},
David A. Golimowski\altaffilmark{\ref{JHU}},
Frederick H. Harris\altaffilmark{\ref{Flagstaff}},
G. S. Hennessy\altaffilmark{\ref{USNO}},
Robert B. Hindsley\altaffilmark{\ref{USNO}},
\v{Z}eljko Ivezi\'{c}\altaffilmark{\ref{Princeton}},
D.Q. Lamb\altaffilmark{\ref{Chicago}},
Jeffrey A. Munn\altaffilmark{\ref{Flagstaff}},
Heidi Jo Newberg\altaffilmark{\ref{Fermilab}},
Ron Rechenmacher\altaffilmark{\ref{Fermilab}},
Donald P. Schneider\altaffilmark{\ref{PennState}}, 
J. Allyn Smith\altaffilmark{\ref{Michigan}},
Chris Stoughton\altaffilmark{\ref{Fermilab}},
Douglas L. Tucker\altaffilmark{\ref{Fermilab}},
Patrick Waddell\altaffilmark{\ref{Washington}}, and
Donald G. York\altaffilmark{\ref{Chicago}} 
}

\newcounter{address}
\setcounter{address}{2}
\altaffiltext{\theaddress}{Princeton University Observatory, Princeton, NJ 08544
\label{Princeton}}
\addtocounter{address}{1}
\altaffiltext{\theaddress}{UKIRT, Joint Astronomy Centre, 660 North
A'ohoku Place, Hilo, HI 96720\label{UKIRT}}
\addtocounter{address}{1}
\altaffiltext{\theaddress}{Gemini North Observatory, 670 North
A'ohoku Place, Hilo, HI 96720\label{Gemini}}
\addtocounter{address}{1}
\altaffiltext{\theaddress}{U.S. Naval Observatory, Flagstaff Station, 
P.O. Box 1149, 
Flagstaff, AZ  86002-1149
\label{Flagstaff}}
\addtocounter{address}{1}
\altaffiltext{\theaddress}{Fermi National Accelerator Laboratory, P.O. Box 500,
Batavia, IL 60510
\label{Fermilab}}
\addtocounter{address}{1}
\altaffiltext{\theaddress}{Apache Point Observatory, P.O. Box 59,
Sunspot, NM 88349-0059
\label{APO}}
\addtocounter{address}{1}
\altaffiltext{\theaddress}{
Department of Physics and Astronomy, The Johns Hopkins University,
   3701 San Martin Drive, Baltimore, MD 21218, USA
\label{JHU}
}
\addtocounter{address}{1}
\altaffiltext{\theaddress}{Department of Physics of Complex Systems,
E\"otv\"os University,
   P\'azm\'any P\'eter s\'et\'any 1/A, Budapest, H-1117, Hungary
\label{Eotvos}
}
\addtocounter{address}{1}
\altaffiltext{\theaddress}{U.S. Naval Observatory, 
3450 Massachusetts Ave., NW, 
Washington, DC  20392-5420
\label{USNO}}
\addtocounter{address}{1}
\altaffiltext{\theaddress}{University of Chicago, Astronomy \& Astrophysics
Center, 5640 S. Ellis Ave., Chicago, IL 60637
\label{Chicago}}
\addtocounter{address}{1}
\altaffiltext{\theaddress}{Department of Astronomy and Astrophysics,
The Pennsylvania State University,
University Park, PA 16802
\label{PennState}}
\addtocounter{address}{1}
\altaffiltext{\theaddress}{University of Michigan, Department of Physics,
	500 East University, Ann Arbor, MI 48109
\label{Michigan}}
\addtocounter{address}{1}
\altaffiltext{\theaddress}{University of Washington, Department of Astronomy,
Box 351580, Seattle, WA 98195
\label{Washington}}
\begin{abstract}
We report the discovery of the coolest field dwarf yet known, selected
as a stellar object with extremely red colors from commissioning
imaging data of the Sloan Digital Sky Survey.  Its spectrum from 0.8
to 2.5$\mu$m is dominated by strong bands of H$_2$O and CH$_4$.  Its
spectrum and colors over this range are very similar to those of
Gliese 229B, the only other known example of a methane dwarf.  It is
roughly 1.2 mag fainter than Gliese 229B, implying that it lies at a
distance of roughly 10 pc. Such a cool object must have a mass well
below the hydrogen-burning limit of 0.08 M$_\odot$, and therefore is a
genuine brown dwarf, with a probable mass in the range 0.015-0.06
M$_\odot$ for an age range of 0.3-5 Gyr.
\end{abstract}

\section{Introduction}
For decades, astronomers have speculated on the nature of substellar
objects or brown dwarfs, objects below the mass necessary to sustain
equilibrium hydrogen thermonuclear burning in their cores (see the
reviews by \cite{Stevenson91} and \cite{Burrows93}).  The past five years
have finally yielded observational evidence for such objects, from deep
near-infrared searches in nearby open clusters (e.g., \cite{Hambly98},
and references therein), in the vicinity of nearby stars
(\cite{Nakajima95}), from proper-motion studies (\cite{Ruiz97}), from
the databases of the Two-Micron All Sky 
Survey (2MASS, \cite{Kirkpatrick99}) and the DENIS survey
(\cite{Tinney97}), and in radial velocity studies of nearby stars (for
a review, see \cite{MarcyButler98}).  The 2MASS survey has defined a
new classification for objects cooler than M stars, referring to these objects
as `L dwarfs'.  These objects have 
surface temperatures low enough (1400-2000 K) that the TiO and VO
bands that dominate the optical spectra of M stars vanish, and
absorption lines of Cs and Rb are seen. 

  There is one object, Gliese 229B, which clearly does not fall into the
L dwarf category.  It was discovered (\cite{Nakajima95}) in an imaging
survey of close companions to nearby young stars (\cite{Nakajima94});
its luminosity and spectrum indicate that it has a temperature of
roughly 900K and a mass of 0.024--0.033 M$_\odot$ for an assumed age of
0.5--1.0 Gyr
(\cite{Leggett99}; see \cite{Marley96} and \cite{Allard96} for earlier
work).  The infrared spectrum of this object 
(\cite{Oppenheimer95}, \cite{Geballe96}, \cite{Noll97}, \cite{Schultz98},
\cite{Oppenheimer98}) is dominated by strong bands 
of H$_2$O, CH$_4$, and CO; CH$_4$ is absent in L dwarfs
(\cite{Noll98}), as it dissociates at temperatures 
above 1300 K (e.g., \cite{Fegley96}, \cite{Burrows97}).  Kirkpatrick
\etal\ (1999) have therefore 
suggested that Gliese 229B be assigned to its own spectral class, T, a class
in which it (until now) has sat alone.  Given that such an object never
reaches a core temperature hot enough to burn hydrogen, the luminosity
and effective temperature are functions of age as well
as mass (see, e.g., Figure 7 of \cite{Burrows97}). 

The near-infrared photometry of Gliese 229B (\cite{Matthews96},
\cite{Golimowski98}, \cite{Leggett99}) shows that it has quite distinctive
near-infrared colors (as predicted by \cite{Burrows97}).  However,
their intrinsically low luminosity (Gliese 229B has a bolometric
luminosity of $6.6 \pm 0.06 \times 10^{-6}\,$L$_\odot$;
\cite{Leggett99}) means that finding further examples will require
deep, wide-angle surveys in the near-infrared.  It is clearly of great
interest to see whether such objects exist in the field, given the
fact that they lie close to the (ill-defined) boundary between planets
and brown dwarfs.

The Sloan Digital Sky Survey (SDSS; 
\cite{GW95}, \cite{SDSS96}, \cite{York99}) is using a dedicated 2.5m
telescope at Apache Point Observatory, New Mexico, 
to obtain CCD images in five broad optical bands
($u'$, $g'$, $r'$, $i'$, $z'$, 
centered at 3540\AA, 4770\AA, 6230\AA, 7630\AA\ and 9130\AA,
\cite{Fukugita96}) over 10,000 deg$^2$ 
of the high Galactic latitude sky centered approximately on the North
Galactic Pole.  Photometric calibration is provided by an auxiliary
$20''$ telescope at the same site.  The survey data processing
software carries out 
astrometric and photometric calibrations, and finds and measures
properties of all objects in the data (\cite{ASTROM}, \cite{Lupton99b}).
The presence of the 
$z'$ band allows the discovery of extremely red objects, which cannot
be  effectively identified in surveys whose red cutoff lies shortward
of 8500\AA.  In particular, 
Fan \etal\ (1999a) have used early SDSS commissioning data to find 15 new 
quasars at $z > 3.65$ (including four with $z > 4.5$), which appear as
red objects due to 
absorbing clouds of intergalactic 
hydrogen, and Fan \etal\ (1999b) have discovered a number of new L dwarfs
from the same dataset.  We here report on follow-up spectroscopy of
an extremely red object in the SDSS imaging data; we
find it to be a near-twin of Gliese 229B, but in the field. 

\section{Observations}

\subsection{SDSS Data}
The equatorial strip in the region of $\alpha = \rm 16^h\,30^m$ was
observed 
twice by the SDSS imaging camera (\cite{Gunn98}), once on 
28 June 1998, with the telescope pointed 2 hours West, and
again on 21 March 1999, with the telescope pointing on the
meridian.  The effective exposure time in each case was 54.1 sec in
each band.  In both cases, the telescope was pointed at the Celestial
Equator, and did not move during these drift-scanning observations.
The seeing in the $z'$ band during these two observations was $1.4''$ and
$1.2''$, respectively.  The object SDSS J162414.37+002915.6
(which we refer to hereafter as \name\ for brevity) was selected for
its extremely red color. 
Tables 1 and 2 give the results of the astrometry and photometry
in these two observations of \name.  It was undetected in $u',
g'$, and $r'$.  Data are quoted as asinh magnitudes (\cite{Lupton99},
\cite{FanQSO}) and are on the AB system (\cite{Fukugita96}).
The $i'$ detection is
at low signal-to-noise ratio, but is consistent between the two
observations.  The $z'$ detection is of very high significance, and
again is consistent between the two observations.  The absolute
calibration of the photometry is uncertain, as the primary standard
star network had not been completely established when these data were
taken; for this reason, we indicate our
photometry with asterisks rather than the primes of the final system,
although we continue to refer to the filters themselves with the prime
notation.
Finding charts for \name\ in the $i'$ and $z'$ bands are shown in
Figure~1.

The $i^* - z^*$
color of $3.77 \pm 0.21$ is unprecedented in the SDSS imaging data taken to date.  
For comparison, M dwarfs with $i^* - z^* > 1.4$ are
quite rare (\cite{FanModel},
\cite{FanQSO}), while L dwarfs get as red as $i^* - z^* \sim 2$
(\cite{Knapp99}).  Gliese 229B has  
not been observed in the SDSS photometric system, but observations in
the Gunn-Thuan system show it to have $i - z = 2.2 \pm 0.3$
(\cite{Nakajima95}).  Synthesizing Gunn-Thuan photometry from our
spectrophotometry of \name\ (see below) gives a color $i - z = 2.4$, in close agreement. 

We have only two observations of \name\ separated by 266 days (the
object is undetected on either POSS-I or POSS-II; \cite{Reid91}), so
we cannot measure both parallax and proper motion from our 
data.  \name\
happens to lie within one of the equatorial astrometric calibration
regions established by Stone (1997), which allowed us to determine the
position relative to this dense grid of stars.  The results are shown
in Table 1, which refers to the positions as measured from the $z'$
detection.  We found that \name\ moved by $-116\pm 31$ mas in right
ascension and by $-27\pm 46$ mas in declination from 1998 June to
1999 March, where the error is the scatter we found for 15 other
stars within $2'$ of \name.  Thus it appears that the position of
\name\ moved significantly in right ascension between the two epochs.
 Assuming a distance of 10 pc (see
below), the change in position due to annual parallax between the two
observations would be $+140$ mas in right ascension and $-40$ mas in
declination.  Thus the detected motion is opposite to the sense in
which annual parallax would move the object.  It is also opposite to
the direction in which uncorrected differential chromatic refraction
would bias the results (given that this object of extreme colors was
observed at different airmasses in the two observations), giving us
some confidence that the motion is both real and is due to proper motion.

\subsection {Optical Spectroscopy}

  We obtained optical spectra of \name\ on
the morning of 20 April 1999 UT using the Double Imaging Spectrograph
on the Apache Point 3.5m telescope, with the same instrumental configuration used by
Fan \etal\ (1999a).  The resolution is 0.0014$\mu$m, and the spectral coverage is
0.4-1.05$\mu$m. 
Observations of the F subdwarf standard BD$\,$+26$\,^\circ2606$
(\cite{OkeGunn83}) provided flux calibration and allowed removal of the 
atmospheric absorption bands.  The seeing was better than $1.2''$ on
this photometric night, and the observations were carried out at low airmass.
The resulting spectrum, a co-addition of three 45-minute exposures, is
shown in Figure~2.  No flux was detected blueward of 0.8$\mu$m,
consistent with the very red $i^* - z^*$ color.  The spectrum shows
a strong H$_2$O absorption band centered at $\sim 0.94\mu$m (which is
robust to the telluric water absorption centered at the same wavelength), the Cs I
line at 0.8523$\mu$m (equivalent width of $12.1 \pm 3.2$\AA) and a strong absorption
line at 1.0017$\mu$m.  We have labeled the latter feature as possible
FeH, although this is a broad feature in L stars, and its strongest
component is over 100\AA\ away, at 0.9896$\mu$m. Moreover, there is a
strong sky line centered at that position.  There is no strong
line at Cs I 0.8943$\mu$m, but 
this region is affected by telluric H$_2$O features.  There is a {\em
possible} detection of a weak band of CH$_4$ in this region of the
spectrum. 

\subsection{Infrared Observations}
Near-infrared photometry (broadband $JHK$) was obtained on 21 April 1999
UT on the United Kingdom Infrared Telescope (UKIRT) using IRCAM, a
camera with a $256\times 256$ InSb array.  The night was photometric, although
the seeing was poor ($1 - 1.5"$).  The data were obtained using the
standard dither technique, and calibrated using UKIRT faint
standards (\cite{Casali92}). The results are shown in
Table 2 where the data are on the UKIRT system; the table also gives
the results of photometry on an AB system, as synthesized from the
spectrum presented below. The colors are almost
identical to those of Gliese 229B (\cite{Leggett99}) but \name\ is 1.2
magnitudes fainter.  There is no evidence that \name\ is extended
beyond the PSF in either the optical or infrared images.

Spectra were obtained in the $J$, $H$, and $K$ bands on the nights of 21 and 22
April and 2 May 1999 UT at UKIRT using the facility grating spectrometer CGS4
(\cite{Mountain90}), which incorporates a $256 \times 256$ InSb array. CGS4
was configured with a $1.2''$ wide slit, 300 mm camera optics,
and a 40 l/mm grating. The spectral region 1.5--1.95$\mu$m was observed
for 800 seconds on 21 April. On 22 April the spectral regions 1.03--1.35,
1.46--2.10, and 1.87--2.51$\mu$m were measured for 1440, 800, and 2560
seconds, respectively, and on 2 May, the 1.19--1.51$\mu$m interval was
observed for 2880 seconds. All spectra were obtained in the standard
stare/nod mode, with the telescope nodded $7.32''$ (12 array rows) between
spectral images. The resolutions in the $J$, $H$, and $K$ bands were
0.0025$\mu$m, 0.0025$\mu$m, and 0.0050$\mu$m, respectively. Wavelength
calibration was accomplished using spectra of krypton, argon, and xenon
lamps and is accurate to better than 0.001$\mu$m. Removal of telluric
and instrumental spectral features and initial flux calibration were
achieved using near simultaneous observations of bright F dwarf stars
(whose strong Brackett and Paschen series hydrogen recombination lines
were removed prior to division) assuming standard visible-infrared
colors. The individual spectra were then combined and scaled so as to
match the near-infrared photometry.

The final, flux-calibrated spectrum, including the initial discovery
spectrum of Figure~2, is shown in
Figure~3. The spectrum looks astonishingly like
that of Gliese 229B, as recalibrated by Leggett \etal\ (1999). In
particular, strong absorption bands of H$_2$O and CH$_4$ dominate the
spectrum, and the individual absorption lines of H$_2$O at 2.0--2.1$\mu$m
discussed
by Geballe \etal\ (1996) are seen as well.  The only significant
difference is the slight excess of flux around 1.7$\mu$m in our
object, and the somewhat stronger lines of K I 1.2432 and 1.2522$\mu$m. 
Note also that while the zero-point of the Gliese 229B spectrum is
slightly uncertain, due to the possibility of miscorrection for scattered
light from Gliese 229A $7''$ away, this is not an issue for our object.
Flux is not detected at the bottom of the H$_2$O band at
1.36--1.40$\mu$m, but is detected in the strongest parts of the H$_2$O bands at
1.15$\mu$m and 1.8--1.9$\mu$m and also at 2.2--2.5$\mu$m.

\section{Discussion}

We have remarked that the colors and spectra of \name\
are quite similar to those of Gliese 229B.  We will assume (although
we cannot demonstrate unequivocally) that \name\ has a similar
effective temperature and luminosity to Gliese 229B (especially given
that the radii of brown dwarfs are almost independent of mass and age;
cf., \cite{Burrows97} and \cite{Burrows99}).  The Hipparcos measured
distance of Gliese 229B is 5.8 pc (\cite{Hipparcos}).  \name\ is roughly
1.2 magnitudes fainter than Gliese 229B in $J$, $H$, and $K$, implying
that it has a distance of 10 pc. 

Objects as cool as \name\ never reach 
equilibrium, and so one cannot infer a mass without
independent constraints on either its age or its surface gravity.  The
surface gravity may be available in the future with more detailed spectral
modeling and higher resolution spectra.  Gliese 229A is classified as
``young disk'' by Leggett (1992), with an inferred age of around
0.6 Gyr, and it is reasonable to assume that it is coeval with Gliese 229B. 
The luminosity and broad band colors of Gliese 229B are consistent with 
models of a 0.5 Gyr-old 0.024 M$_\odot$ object.  We have no
direct measurement of the age of \name, except to say that it is
not obviously associated with a star-forming region.  Assuming the
temperature and luminosity are similar to those of  Gliese 229B,
the mass of \name\ probably lies in the range 0.015--0.06 M$_\odot$ for an age 
range of 0.3--5 Gyr, based on a comparison to models by
Burrows \etal\ (1997).

Assuming a distance of 10 pc, the change of position of \name\ in our
two observations implies an entirely plausible transverse velocity of
$17 \pm 4\,\rm km\,s^{-1}$.  However, we note that the star $\sigma$
Ser, at a distance of 28 pc (\cite{Hipparcos}), lies 
only 0.76$^\circ$ away from \name, and shares its proper motion within
1.5$\sigma$.  If \name\ were at the
same distance, then these two stars could be a wide binary, separated
by 0.4 pc. However, this would require a luminosity for \name\ of
$10^{-4.3}\,L_\odot$, which would require an
implausibly low age and high temperature (approaching that at which
CH$_4$ can no longer exist) from the Burrows \etal\ (1997)
models.  An accurate parallax will of course tell us whether this
should be taken seriously. 

\name\ is the reddest object found in the SDSS database from roughly
400 square degrees of imaging data, or roughly 1\% of the celestial
sphere.  If this region of sky is typical, there should be of order
100 comparable objects in the sky, and the SDSS in 
particular will discover of order 25 of them (as it will survey
1/4 of the celestial sphere).  Indeed, 400 square degrees may be an
overestimate of the effective area from which \name\ was selected, as
not all of the area surveyed was observed in optimal seeing, and we used the
fact that we obtained consistent photometry of \name\ on two separate
observations to bolster our confidence that the photometry was correct.
In any case, assuming that \name\ is at 10 pc, and recognizing the
dangers of statistical arguments based on a single object, we can
infer a volume density of 0.03 objects per cubic parsec, which would
imply that the nearest of these objects is less than 4 pc away (and
therefore more than 2 magnitudes brighter than \name!).  There is not
yet a careful measurement of the space density of L dwarfs, but based
on the statement in Kirkpatrick \etal\ (1999) that 6/25 L dwarfs are within
25 pc, one infers a lower limit to their space density of 0.01 objects
per cubic parsec.  Therefore, the data are consistent with roughly
comparable space densities of L dwarfs and methane dwarfs.

  The SDSS is not sensitive to objects of this temperature that are
substantially fainter than $z^* = 19$.  With an $i^* - z^*$ of 3.5, we
quickly reach our photometric limits of $i^* \approx 22.5$, $z^*
\approx 20.8$ for 5$\sigma$ detections of stellar
sources in $1''$ seeing (\cite{Gunn98}).  However, the combination of $i'$ and
$z'$ photometry from the SDSS, and $JHK$ photometry from the 2MASS
survey, will be particularly powerful for finding such objects. 

The Sloan Digital Sky Survey (SDSS) is a joint project of the
University of Chicago, Fermilab, the Institute for Advanced Study, the
Japan Participation Group, The Johns Hopkins University, the
Max-Planck-Institute for Astronomy, Princeton University, the United
States Naval Observatory, and the University of Washington.  Apache
Point Observatory, site of the SDSS, is operated by the Astrophysical
Research Consortium.  Funding for the project has been provided by the
Alfred P. Sloan Foundation, the SDSS member institutions, the National
Aeronautics and Space Administration, the National Science Foundation,
the U.S. Department of Energy, and the Ministry of Education of Japan.
MAS and XF acknowledge additional support from Research Corporation,
NSF grant AST96-16901, the Princeton University Research Board, and 
an Advisory Council Scholarship, and GK is grateful for support from
Princeton University and NSF grant AST96-18503.  
We also thank Russet
McMillan for her usual expert assistance at Apache Point Observatory,
Jen Adelman for helping on the data reduction, 
and Davy Kirkpatrick, Herbert Strauss, and Scott
Tremaine for some
very enlightening discussions.  UKIRT is operated
by the Joint Astronomy Centre on behalf of the U.K. Particle Physics
and Astronomy Research Council. We are grateful to the staff of UKIRT
for its support, and to A. J. Adamson for use of UKIRT Director's time
and for obtaining some of the CGS4 data.

\newpage
\begin{center}
Table 1. Optical Positions and SDSS Photometry of Methane Dwarf

\begin{footnotesize}
\begin{tabular}{lcccccc}\\ \hline \hline
Position & $u^*$ & $g^*$ & $r^*$ & $i^*$ & $z^*$ &Date \\ \hline
16:24:14.37$\,$+00:29:15.8 & $25.07 \pm 0.39$ & $25.95 \pm 0.35$ &
$25.34 \pm 0.56$ & $22.70 \pm 0.27$ & $19.02 \pm 0.04$ & June 1998 \\
16:24:14.36$\,$+00:29:15.7 & $24.29 \pm 0.33$ & $24.29 \pm 0.39$ &
$24.18 \pm 0.53$ & $22.88 \pm 0.32$ & $19.03 \pm 0.04$ & March 1999 \\
\hline \hline
\end{tabular}
\end{footnotesize}
\end{center}

\noindent Astrometry is given in J2000 coordinates. Photometry is
reported in terms of asinh magnitudes, which 
smoothly goes to a linear scale in the limits of low signal-to-noise
ratio; see
Lupton, Gunn \& Szalay (1999) and Fan \etal\ (1999a) for details.  The $u^*$, $g^*$, and
$r^*$ values all represent non-detections (for comparison, in our
system, zero flux corresponds to 24.24, 24.91, 24.53, 23.89, and
22.47, in $u^*$, $g^*$, $r^*$, $i^*$, and $z^*$, respectively). 
The definition of the
photometric system is still uncertain at the level of roughly 0.05 mag in all
bands; we quote measured values using asterisks (to
represent preliminary photometry) rather than the primes of the final
system. 

\bigskip\bigskip

\begin{center}
Table 2. Near-Infrared Photometry of Methane Dwarf

\begin{footnotesize}
\begin{tabular}{cccccl}\\ \hline \hline
$i$ & $z$ & $J$ & $H$ & $K$ \\ \hline
21.2 & 18.79 & $15.53 \pm 0.03$ & $15.57 \pm 0.05$ & $15.70 \pm 0.05$
& Vega system\\
21.6 & 19.21 & $16.44 \pm 0.03$ & $16.95 \pm 0.05$ & $17.56 \pm 0.05$
& AB system\\
\hline \hline
\end{tabular}
\end{footnotesize}
\end{center}

\noindent The $JHK$ photometry on the first line is on the UKIRT
system, which references colors to the color of Vega.  The
$i$ and $z$ photometry is on the Gunn-Thuan system; it, and the $JHK$
AB photometry, are synthesized from the composite spectrum of
Figure~3. 
\newpage
\begin{figure}
\vspace{-0.5cm}
\epsfysize=600pt \epsfbox{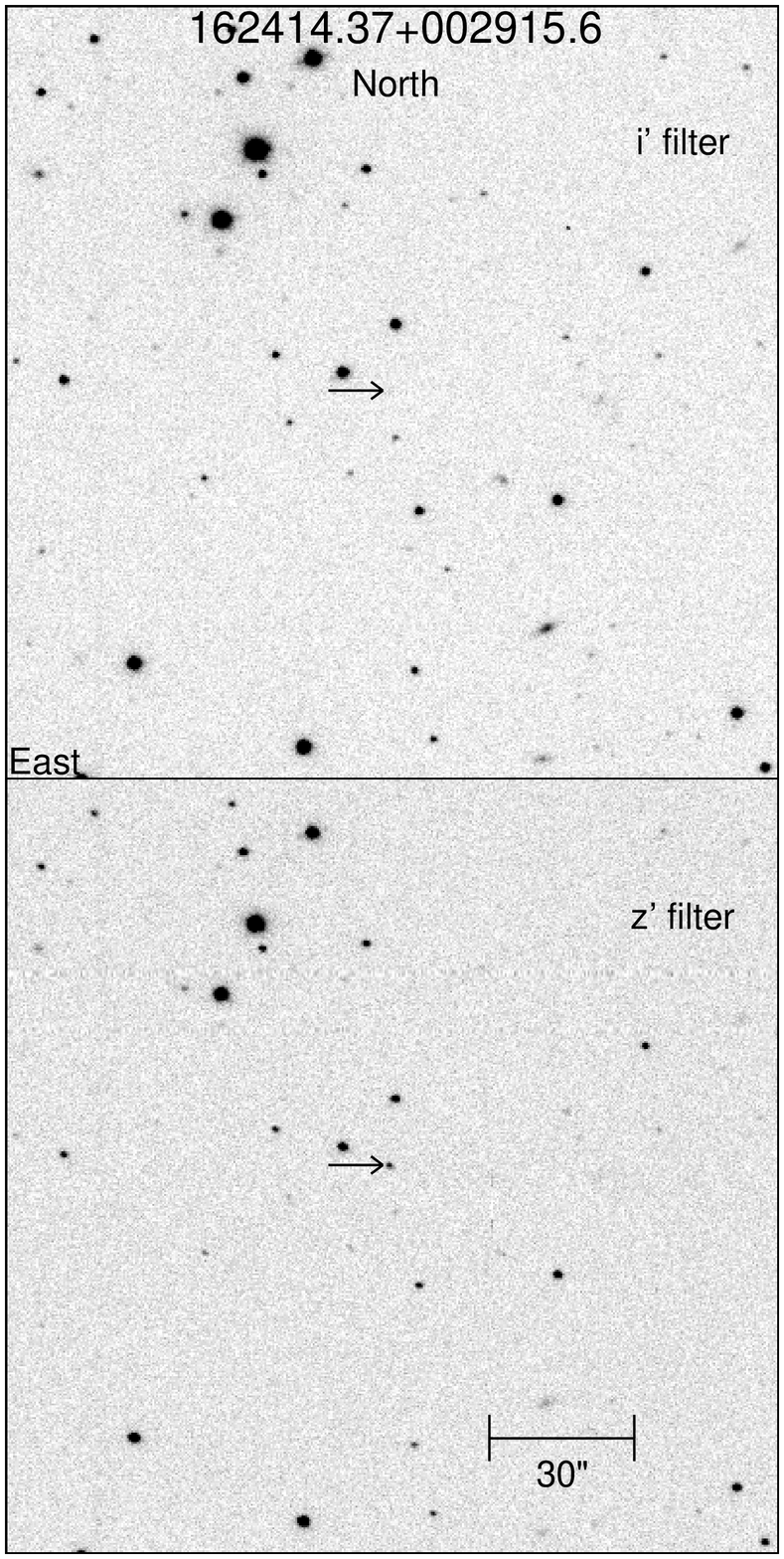}
\vspace{0.5cm}
Figure 1. Finding chart for \name\ (discovery image from the
SDSS).  The field is $160''$ on a 
side.  The field is given in both the $i'$ and $z'$ bands (54.1 sec
exposure time) from data taken on 21 March 1999. 
North is up; East is to the left. 
\label{fig:finding-chart}
\end{figure}
\newpage

\begin{figure}
\vspace{-0.5cm}
\epsfysize=500pt
\vspace{0.5cm}\epsfbox{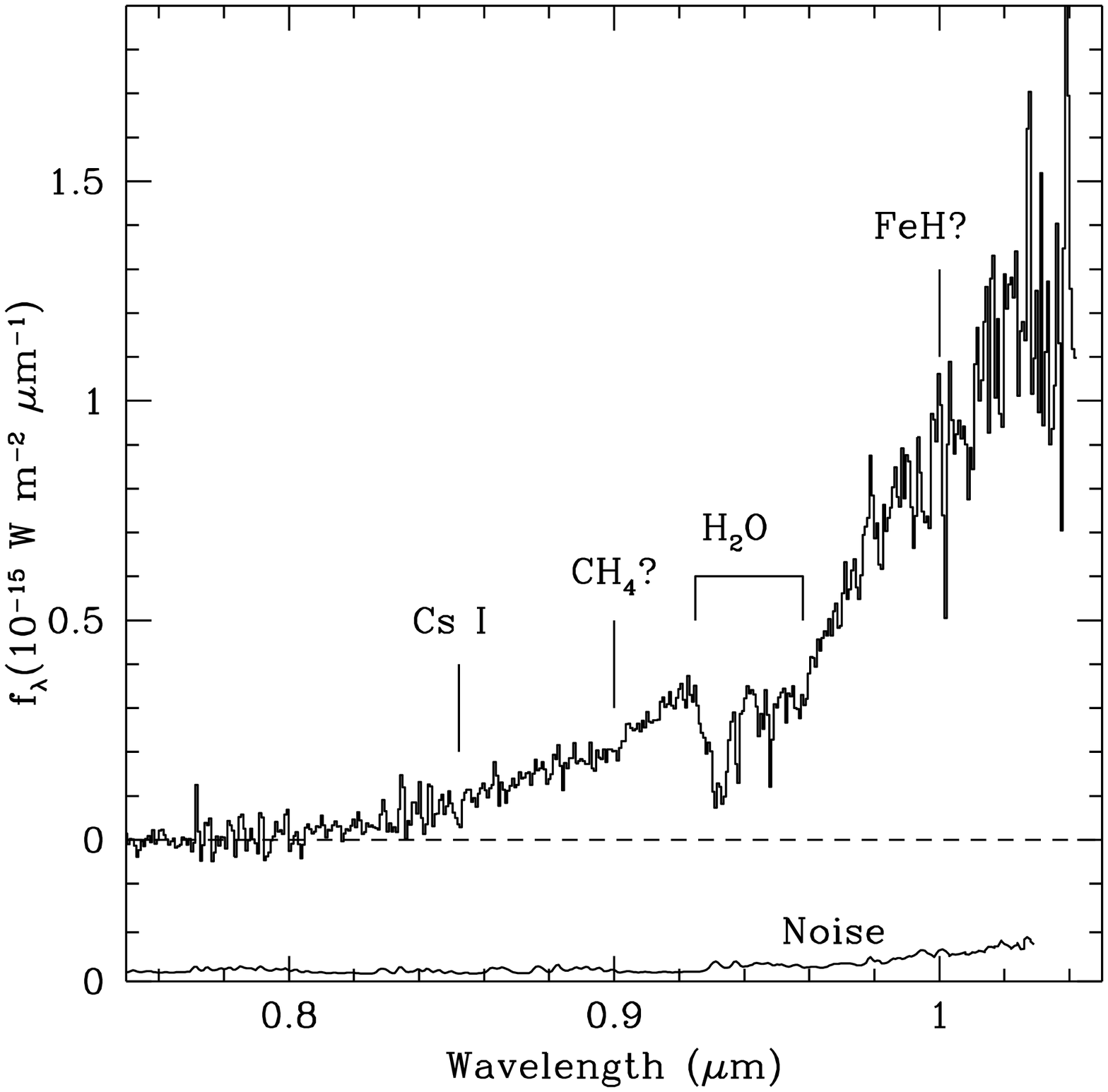}
Figure 2. The discovery optical spectrum of \name, from the DIS
spectrograph on the Apache Point 3.5m, at 0.0014$\mu$m resolution.
The estimated noise in this spectrum is given as well.  Observed
features are marked.  
\label{fig:optical-spectrum}
\end{figure}

\begin{figure}
\vspace{-0.5cm}
\epsfysize=500pt \epsfbox{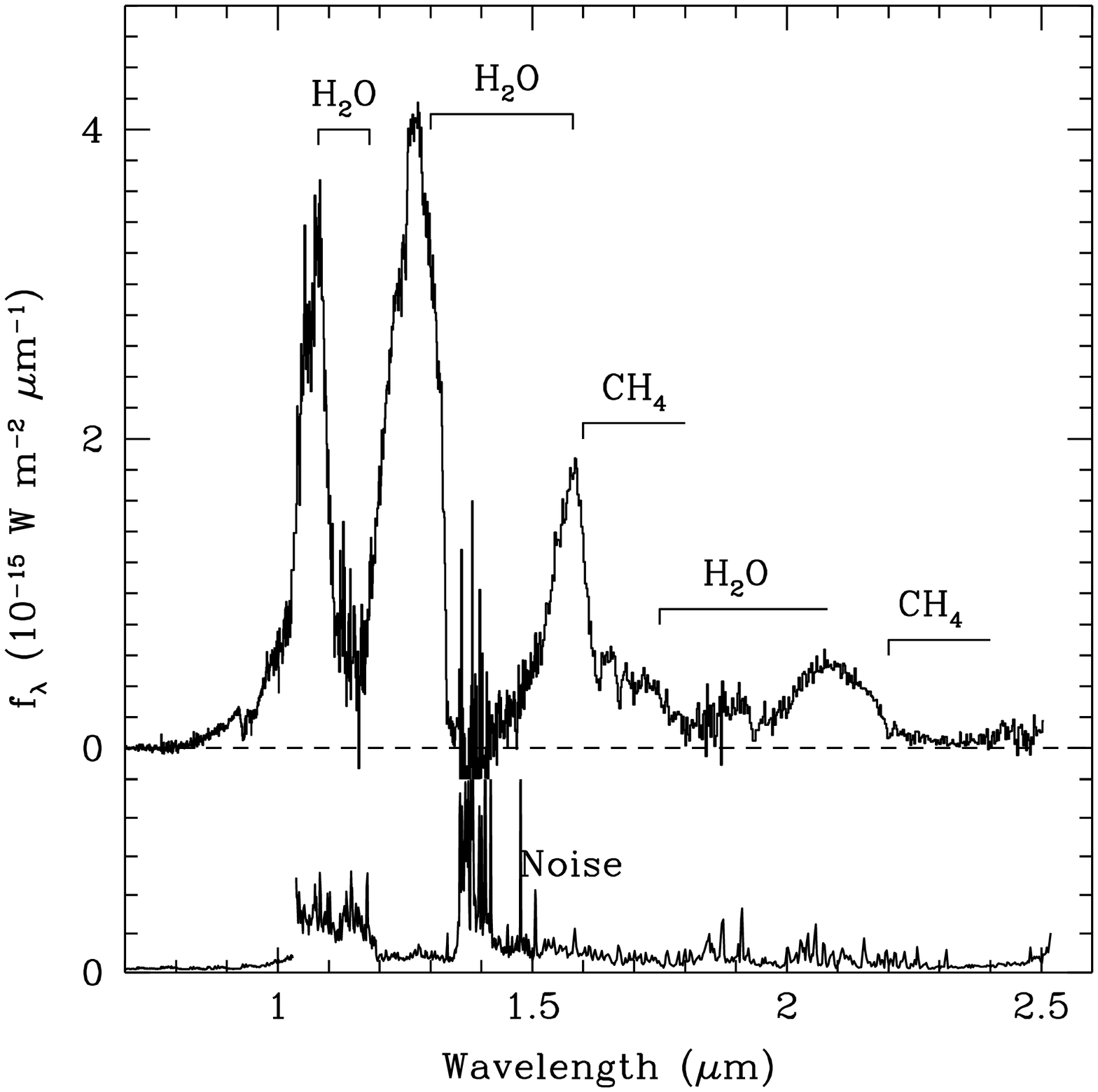}
\vspace{0.5cm}
Figure 3. The combined optical and $JHK$ spectrum of \name; the latter
was taken with the CGS4 at UKIRT, with 0.0025--0.0050$\mu$m resolution.  The
estimated noise in this spectrum is given as well.  The prominent
bands of H$_2$O and CH$_4$ are marked.  Most of the narrow spectral
features at 1.2--1.3, 1.5--1.7, and 1.95--2.1$\mu$m are real. 
\label{fig:full-spectrum}
\end{figure}
\newpage

\end{document}